\documentstyle[aps,prl,preprint]{revtex}              
\draft
\title{Low frequency admittance of a quantum point
contact}
\author{T. Christen and M. B\"uttiker}
\address{D\'epartement de physique th\'eorique,
Universit\'e de Gen\`eve, 24 Quai Ernest-Ansermet \\ 
CH-1211 Gen\`eve, Switzerland}
\begin{document}
\maketitle
\begin{abstract}
\end{abstract}
We present a current and charge conserving theory 
for the low frequency admittance of a quantum point contact.
We derive expressions for the
electrochemical capacitance and the displacement current.
The latter is determined by the {\em emittance} which equals the
capacitance only in the limit of vanishing transmission.
With the opening of channels the capacitance and the emittance decrease
in a step-like manner in synchronism with the conductance steps. For
vanishing reflection, the capacitance vanishes and the emittance is
negative. 

\indent
There is growing interest in transport properties of electric
nanostructures such as quantum point contacts, quantum
wires, and quantum dots, to mention but a few \cite{rev,IMRY}.
These mesoscopic conductors can be so small that transport at low
temperatures is phase coherent or even mainly ballistic including
only a few elastic scattering events. 
The scattering approach to electrical conduction \cite{IMRY,Bu86,LAND}
has successfully been used to describe many experiments.
For a phase coherent conductor with two probes this theory relates
the transmission probabilities $ T^{(j)}$ of the
occupied one-dimensional subbands to the dc-conductance
$G^{(0)}=(2e^{2}/h) \sum T^{(j)}$. The validity of this 
conductance formula was experimentally confirmed first by van Wees
et al. \cite{QPCv} and Wharam et al. \cite{QPCW} who found a stepwise
increase of the conductance by successively opening conduction channels
of a quantum point contact.\\ \indent
A more novel concept concerns the notion of the {\em mesoscopic
capacitance}. Usually, the capacitance $C_{\mu}$ is defined by the
static change of the charge
on a conductor as a response to the electrochemical voltage drop between
the contacts. However, there exists also a dynamic point of
view which is important for practical use: the capacitance
is then associated with the phase shift between a current and a voltage
oscillation at small frequencies $\omega $, i.e. with the imaginary
part of the low frequency admittance $G(\omega)$ of a resistor and condenser
in parallel. A dynamical derivation of a mesoscopic capacitance was
given by B\"uttiker, Thomas, and Pr\^etre \cite{BTPC}. To make a clear
distinction between the static and the dynamic conceptions, we call
$E_{\mu} = i(dG/d\omega)_{\omega =0}$ the {\em emittance} of a
conductor. For a purely capacitive structure such
as a condenser the static and
dynamical derivations lead to identical results, i.e.
$E_{\mu}= C_{\mu}$. This case is characterized by a displacement
current entering the sample through the leads which is equal to
the change of the charge on a condenser plate. We mention that
in a mesoscopic sample the relevant density of states
(DOS), $dN_{1}/dE$ and $dN_{2}/dE$, of the `mesoscopic condenser plates'
can be so small that $C_{\mu}$ is no longer equal to the
geometric capacitance $C_{0}$ but depends on the
DOS \cite{BTPC}: $C_{\mu}^{-1}= C_{0}^{-1} +D_{1}^{-1} +D_{2}^{-1} $
with $D_{k}=e^{2}dN_{k}/dE$. This is due
to the fact that the voltage drop between the reservoirs can differ
significantly from the drop of the electrostatic potential at the
plates. On the other hand, for conductors
which permit transmission $E_{\mu}= C_{\mu}$ is {\em not}
valid. In this Letter, we derive expressions for the capacitance and
the emittance of a quantum point contact (Fig. \ref{fig1}).
The model which we develop
also describes a ballistic quantum wire containing an elastic scatterer
or a mesoscopic condenser with tunneling between the two condenser
plates (leakage). Capacitance properties of small conductors with
transmission are also of great importance in
tunneling microscopy \cite{STM}.\\ \indent
First, we present our results for a single-channel conductor.
Subsequently, their derivation and the generalization to many channels 
is provided using the scattering approach to
low-frequency transport developed in Refs. \cite{BTPE1,BuetE}.\\ \indent
Consider a phase-coherent single-channel conductor containing
a localized scattering obstacle. It turns out that $C_{\mu}$ and
$E_{\mu}$ decrease for increasing transmission probability $T=1-R$.
In particular, we show that the capacitance is proportional to the
reflection probability $R$
\begin{equation}
C_{\mu} = \frac{R}{C_{0}^{-1} +D_{1}^{-1} +D_{2}^{-1} } \;\;.
\label{cmu}
\end{equation}
In general, also the geometric capacitance $C_{0}$ depends
on $R$. For example, $C_{0}^{-1} $ decreases
for two condenser plates approaching each other.
However, since the $D_{k}$ are nearly independent of $T$
and remain finite for $R\to 0$ one concludes from Eq. (\ref{cmu})
that $C_{\mu}$ vanishes for $R\to 0$ even if $C_{0}^{-1}$ vanishes.
This is reasonable since for ideal transmission (no barrier)  
a charge accumulation does not occur. For $R=1$,
on the other hand, we recover from Eq. (\ref{cmu}) the above
mentioned expression for the electrochemical capacitance of a
mesoscopic condenser.\\ \indent
The emittance of a single-channel conductor is given by
\begin{equation}
E _{\mu} = C_{\mu}R -\frac{D}{4}T^{2} \;\;,
\label{E}
\indent
\end{equation}
where $D=D_{1}+D_{2}$ is associated with the total (relevant)
DOS of the sample. As expected, $R=1$
implies $E_{\mu}=C_{\mu}$. On the other hand, for total transmission
($R=0$) the emittance is negative, $E_{\mu}=- D/4$. 
For the particular case where the geometric capacitance is
sufficiently large and where the sample is spatially symmetric,
i.e. $C_{0} \gg D_{1} = D_{2}$, we find $E_{\mu} =
(D/4)\:(R-T)$. This illustrates a cross-over between
positive and negative emittance. Negative emittances are
characteristic for conductors with nearly perfect transmission.
For resonant tunnel junctions an inductive-like kinetic response 
is discussed in Refs. \cite{PRICE,FUDU,FREN}. In Ref. \cite{BTPE1} it
is shown that the emittance remains negative even when the charge in
the well is totally screened. It is interesting that
the emittance for the symmetric tunnel resonance barrier in this limit
can also be written as $E_{\mu} = (D/4)\:(R-T)$.
A similar relation has been found by Mikhailov and Volkov
\cite{MiVo} who calculated with a Boltzmann approach the low
frequency plasma-wave spectrum for a tunnel junction. Introducing a
time $\tau _{T}$, they found a tunnel contribution $C_{T}$
to the capacitance which is proportional to $\tau _{T} (R-T)$.
Although their result is not in full accordance with Eq. (\ref{E}),
it holds $E_{\mu} = C_{T}$ if the barrier is symmetric,
if capacitances in series and in parallel are neglected,
and if one replaces $\tau _{T}$ by the expression $hD/(2e^{2})$.
Furthermore, we showed in Ref. \cite{ChBu} that positive and negative
emittances exist in quantized Hall samples, depending on whether
edge states provide perfect transmission or perfect reflection
channels.\\ \indent
Consider now a quantum point contact (Fig. \ref{fig1})
connected on either side to reservoirs $\alpha $ ($=1,2$).
A variation of the voltage
$\delta V_{\alpha}=\delta \mu _{\alpha}/e$
in reservoir $\alpha $ changes the electrochemical potential
$\delta \mu _{\alpha}$ of the incoming particles which are
partly scattered back and partly transmitted. The admittance matrix
$G_{\alpha \beta} (\omega )= \delta I_{\alpha} 
/ \delta V_{\beta} $ represents
the linear response of the current $\delta I_{\alpha} $
through contact $\alpha $ for a small voltage oscillation
$\delta V_{\beta}  \propto \exp (-i\omega t)$ in reservoir ${\beta}$.
For low frequencies one can write 
\begin{equation}
G_{\alpha \beta} (\omega ) = G_{\alpha \beta }^{(0)} - i \omega
E _{\alpha \beta} \;\;,
\label{G}
\end{equation}
where $E_{\alpha \beta}$ is the emittance matrix.
A microscopic calculation of the emittance is a complicated task 
since the electrostatic potential is a complicated function of
space. The aim of this work is to develop a simple model that
captures the essential physical features.\\ \indent
First, we mention that an applied voltage can polarize the
conductor but leaves the total charge unaffected. Hence,
for a conductor in electrical isolation (no other nearby conductors
or gates) charge and current are conserved, meaning 
$G_{11}=G_{22}=- G_{12}= -G_{21}$ $ \equiv G \equiv G^{(0)} -i\omega
E_{\mu }$. The non-equilibrium charge distribution with the
form of a dipole has a charge $\delta q_{1}$ to the left
and a charge $\delta q_{2} = -\delta q_{1}$ to the right
of the barrier. Instead of treating the entire potential landscape
realistically we introduce only two potentials $ \delta U_{1,2}$
for the regions $\Omega_{1,2}$ (dark regions in Fig. \ref{fig1}).
These regions are characterized by an incomplete screening of the
excess charge. Consider for a moment a voltage shift $\delta V_{1}=\delta
\mu_{1}/e$ only in the left reservoir. On the far left
side of the point contact one has complete screening, thus the local
electric potential shift follows the electrochemical potential,
$\delta \mu_{1}/e $, while on the
far right side it vanishes. As we move along the conductor
from the left reservoir to the right reservoir the potential shift 
drops from $\delta \mu_{1}/e$ to $\delta U_{1}$ to $\delta U_{2}$
to zero. The potential drop will be strongly
localized near the maximum of the barrier in the center
of the quantum point contact. In fact, the potential drop will be
localized within a screening length. We discretize this
potential \cite{PBT}. We emphasize that within the framework of the 
general approach provided by  Ref. \cite{BuetE} the complicated
full quantum mechanical and space dependent problem
can be treated analogously.\\ \indent
In the basis of eigen-channels the transmission problem
through a quantum point contact can be represented as a sum of
single-channel transmission-problems \cite{GLAZ,BU90}.
The potential of a quantum point contact has the shape of a 
saddle \cite{BU90} with a value $eU_{0}$ at the saddle point.
Near the saddle the potential can also be separated into a longitudinal
part $eU(x)$ and a transverse part $eU(y)$.
Thus, in a first step we consider a single-channel transmission problem
in a potential $eU(x).$
The variation of this potential is slow compared to the Fermi wavelength
which allows us to use the semiclassical WKB approximation
for the local density of states $dn(x)/dE$ and for the transmission
probability $T$ \cite{LaLi,MiGo}. The regions $\Omega_{k}$ to the left and
to the right of the barrier in which the potentials are not screened are
$\Omega_{1}= [-l_{1}, -x_{1}]$ and $\Omega_{2}= [x_{2}, l_{2}]$,
respectively, where the $l_{1,2}$ are of the order of the screening length.
The $x_{k}$ are determined by the WKB turning points if 
$E_{F}< eU_{0}$, and they are given by $x_{k}=0$ (the location of
the barrier peak) for $E_{F}\geq  eU_{0}$. We express the DOS
in the region $\Omega_{k}$ in the form of a quantum capacitance 
\begin{equation}
D_{k} = e^{2} \int _{\Omega_{k}} dx \; \frac{dn(x)}{dE}\;\;.
\label{Dk}
\end{equation}  
\indent
For the following we need the nonequilibrium state, i.e. the charge 
$\delta q_{k}$ which resides in $\Omega_{k}$ as a
consequence of a voltage variation $\delta V _{\alpha} =
\delta \mu_{\alpha}/e$ at contact $\alpha$. This charge can be found
with the help of the {\em partial densities of states}
(PDOS) associated with
carriers in $\Omega_{k}$  scattered from contact $\beta $ to contact
$\alpha $ \cite{BuetE}. In the semiclassical (WKB) case this PDOS
is given by
\begin{equation}
D_{\alpha k \beta} =
D_{k} \left( T/2 + \delta _{\alpha \beta}(R\: \delta _{\alpha
k}-T/2)\:\right) \;\;,
\label{pdos}
\end{equation}
where $\delta _{kl}$ is the Kronecker delta. Note that
$D_{k}=\sum _{\alpha \beta} D_{\alpha k \beta}$. Greek and roman indices 
label reservoirs and incompletely screened regions near the point
contact, respectively. The injected charges lead to induced electrostatic
potentials $\delta U_{k}$ which counteract the built up of charge
in the regions $\Omega _{k}$, i.e. the shifts $\delta U_{k}$ of the
band bottoms induce a charge response. For a spatially slowly varying
potential this response is local and is determined
by the DOS, $\delta q^{ind}_{k} = -D_{k} \delta U_{k}$.
The charge in region $k$ is then given by 
\begin{equation}
\delta q_{k}=\sum _{\alpha \beta}D_{\alpha k \beta}(\delta V
_{\beta}-\delta U_{k}) \equiv \sum _{\beta} \overline{D}_{k
\beta}(\delta V_{\beta}-\delta U_{k}) \;\;,
\label{dqk} 
\end{equation}
were we introduced the {\em injectivity} \cite{BuetE}
$\overline{D}_{k \beta}=\sum _{\alpha} D_{\alpha k \beta} $ which is 
the PDOS of region $\Omega_{k}$ associated with carriers injected
at contact $\beta $.\\ \indent
We next determine the electrochemical capacitance.
We introduce a geometrical capacitance matrix
$C_{0,kl}=(-1)^{k+l}C_{0}$ associated with the regions $\Omega _{k}$,
which we assume to be known. 
In general, it is found by solving the Poisson equation. 
The electrostatic and electrochemical capacitance matrices 
$C_{0,kl}$ and $C_{\mu , k\beta}$, respectively, relate the charge
to the potentials via
\begin{equation}
\delta q _{k} = \sum _{l} C_{0 , k l} \; \delta U_{l}
= \sum _{\beta} C_{\mu , k \beta} \; \delta V_{\beta} \;\;.
\label{cap}
\end{equation}
Charge conservation implies $C_{\mu,k \beta}=(-1)^{k+\beta}C_{\mu}$. 
Using Eqs. (\ref{pdos})-(\ref{cap}) yields then the electrochemical
capacitance (\ref{cmu}).\\ \indent
To calculate the emittance matrix
we remark that $E_{\alpha \beta}\: \delta V_{\beta}$ corresponds to the
displacement charge $\delta Q_{\alpha}$ which passes contact $\alpha $
for a variation $\delta V _{\beta}$ 
of the voltage in reservoir $\beta $. Note that
$\delta q_{k} = \delta Q_{\alpha=k} $ is only valid
if $R=1$ but does not hold if $R <1$. 
Since we restrict ourselves to the first-order frequency term,
it is sufficient to calculate
the quasi-static displacement charge. We take the Coulomb interaction
into account self-consistently by considering two contributions to
$\delta Q_{\alpha}$. A first part which neglects
screening is given by the kinetic contribution $ D_{\alpha \beta}
\delta V_{\beta} $, where $D_{\alpha \beta}= \sum _{k} D_{\alpha
k \beta} $ is the PDOS of carriers scattered from contact $\beta$
to contact $\alpha $ at fixed electrostatic potentials. 
A second part corresponds to a screening charge which is due to the
shifts $\delta U_{k}$ of the band bottoms. The part of the screening charge
which is eventually scattered to contact $\alpha $ is then
given by $ -\sum _{k \gamma} D_{\alpha k \gamma} \delta U_{k}\equiv $
$-\sum _{k} \underline{D}
_{\alpha k} u_{k \beta}\: \delta V_{\beta}$, where we defined
the {\em emissivity} \cite{BuetE} 
$\underline{D} _{\alpha k} = \sum_{\gamma} D_{\alpha k \gamma}$
associated with the states scattered from
the region $\Omega _{k}$ to contact $\alpha $. Furthermore, we introduced
the {\em characteristic potentials} \cite{BuetE}
$u_{k \beta} = \partial U _{k}/ \partial
V_{\beta}$ which give the change of the electrostatic potential
in region $k$ due to a variation of the voltage in reservoir
$\beta $. The negative sign of the screening charge is due to
the fact, that a positive shift of the band bottom at fixed
electrochemical potential diminishes the number of charge carriers.
One finds from Eqs. (\ref{dqk}) and (\ref{cap})
$u_{k \beta}= (\overline{D}_{k \beta}-c_{\mu, k \beta}
)/D_{k}$. 
The emittance matrix is the sum of kinetic and screening
charges scattered to contact $\alpha $ \cite{BuetE}
\begin{equation}
E_{\alpha \beta }= D_{\alpha \beta }- \sum _{k}
\underline{D}_{\alpha k} u_{k \beta}  \;\;.
\label{Eab}
\end{equation}
Using the total density of states $D = D_{1}+D_{2} = \sum_{\alpha k}
\underline{D}_{\alpha k} = \sum_{\alpha \beta} D_{\alpha \beta}$
of both regions $\Omega_{1}$ and $\Omega_{2}$, the expression
(\ref{pdos}) for the PDOS, and the characteristic potentials
given above, we find Eq. (\ref{E}) for the emittance of a single-channel
mesoscopic conductor.\\ \indent
In order to generalize the results (\ref{cmu}) and (\ref{E})
to $M$ channels $j= 1,...,M$ with channel thresholds $E^{(j)}_{b}$
we use the fact that the total PDOS is the sum of the PDOS of the
single channels, i.e. $D_{\alpha k \beta} = \sum _{j}
D_{\alpha k \beta}^{(j)}$. If $E_{F}< E_{b}^{(j)}$,
the PDOS for the channel $j$ vanish, 
$D_{\alpha k \beta}^{(j)}(E_{F}) \equiv 0$. If $E_{F} \geq E_{b}^{(j)}$,
the PDOS $D_{\alpha k \beta}^{(j)}(E_{F})$ are given by the single-channel
PDOS (\ref{pdos}) taken at an energy $E_{F}-E_{b}^{(j)}$.
Proceeding the same way as above, we find an electrochemical capacitance
(\ref{cmu}) with a reflection probability
$R=1-T_{1}/2-T_{2}/2$, where $T_{k}= D_{k}^{-1}\sum _{j}
T^{(j)}D_{k}^{(j)}$ is an average transmission probability weighted   
by the density of states of $\Omega _{k}$. For the emittance we find
\begin{equation}
E_{\mu} = C_{\mu}R -\frac{1}{4}(D_{1}T_{1}^{2}+D_{2}T_{2}^{2}) \;\;.
\label{Ej}
\indent
\end{equation}
Equation (\ref{Ej}) applies to a scattering obstacle in an N-channel wire.
Let us now apply this result to the quantum point contact of Fig.
\ref{fig1}. One expects a step-like behavior of the capacitance and
the emittance
as the number of open channels increases. In the following
we consider a symmetric barrier with the quadratic potential
$U(x)=U_{0}(b^{2}-x^{2})/b^{2}$ if $|x|\leq b$,
and $U(x)=0$ if $b<|x|\leq l$. For this special case, the PDOS and
the transmission probability can be calculated analytically
from the WKB expressions \cite{LaLi,MiGo}. For simplicity, we assume a
constant electrostatic capacitance $C_{0}= 1\: fF$ between $\Omega
_{1}$ and $\Omega _{2}$ and a fixed number
of occupied channels in these regions. The only parameter to be varied is
the potential height $U_{0}$. We assume that no additional
channels enter into the regions $\Omega _{k}$ during the variation of
$U_{0}$. In Fig. \ref{fig2} we show the result
for a constriction with $b=500\: nm$, $l = 550 \: nm$, and
with three equidistant channels separated by $E_{F}/3=7/3 \: meV$.
The dotted, dashed, and solid curves correspond to
the dc-conductance, the electrochemical capacitance, and the
emittance, respectively. For small $U_{0}$ where all channels are
open, the capacitance vanishes and the emittance is 
negative. At each conductance step, the capacitance and the emittance
increase and eventually merge when all channels are closed.
Due to a weak logarithmic divergence of the WKB density of
states at particle energies $E=eU_{0}$ (where WKB is
not appropriate), the WKB emittance diverges weakly
(steep edges of the emittance below and above the steps). 
A more accurate quantum mechanical calculation of the PDOS from
the scattering matrix \cite{BTPE1,BuetE} yields a suppression
of these divergencies, i.e. the regions between the steps
become more flat.\\ \indent
To summarize, we present a theory for the capacitance and the
low frequency admittance of one-dimensional mesoscopic two-terminal
conductors. The electrochemical capacitance
defined as the charge variation on a conductor for a voltage drop
in the reservoirs turns out to be proportional to
the reflection probability. The quantity
which is usually measured in an experiment is not the charge (or the
dipole moment) but rather the displacement current which is determined
by the emittance. The emittance equals the capacitance
for vanishing transmission but becomes negative if transmission
predominates. Quantum point contacts which provide the possibility
to vary the number of open one-dimensional channels should thus
show not only conductance steps but also steps in
the capacitance and the emittance. The generalization to
conductors which are not in electrical
isolation will be published elsewhere.
We only mention that metallic gates used to form
the point contact couple with a purely capacitive emittance which
exhibits peaks as new channels are opened.   
Furthermore, the presence of gates causes the zero in the
emittance of the point contact to be shifted to larger values of $T$ ($<1$).
We believe that
the presented theory is also a starting point in order to treat the
finite-frequency noise of quantum point contacts including
Coulomb interactions \cite{SN}.\\ \indent
{\em Acknowledgement}
This work has been supported by the Swiss National Science
Foundation. 

\newpage

\begin{figure}
\caption{Quantum point contact connected to reservoirs 
with electrochemical potentials $\mu _{\alpha } =
\mu _{0}+\delta \mu
_{ \alpha} $, and for the particular case of one transmitted and
two backscattered channels inside $\Omega _{k}$
(dark regions) with electric potentials $\delta U_{k}$.}
\label{fig1}
\end{figure}
\begin{figure}
\caption{Dependence of the conductance (in units $2e^{2}/h$; dotted
curve), capacitance and emittance (in units of $fF$; dashed and full
curves, respectively) on the barrier height $eU_{0}$ 
for a quantum point contact with three relevant channels (see Fig.
\protect\ref{fig1}).}
\label{fig2}
\end{figure}

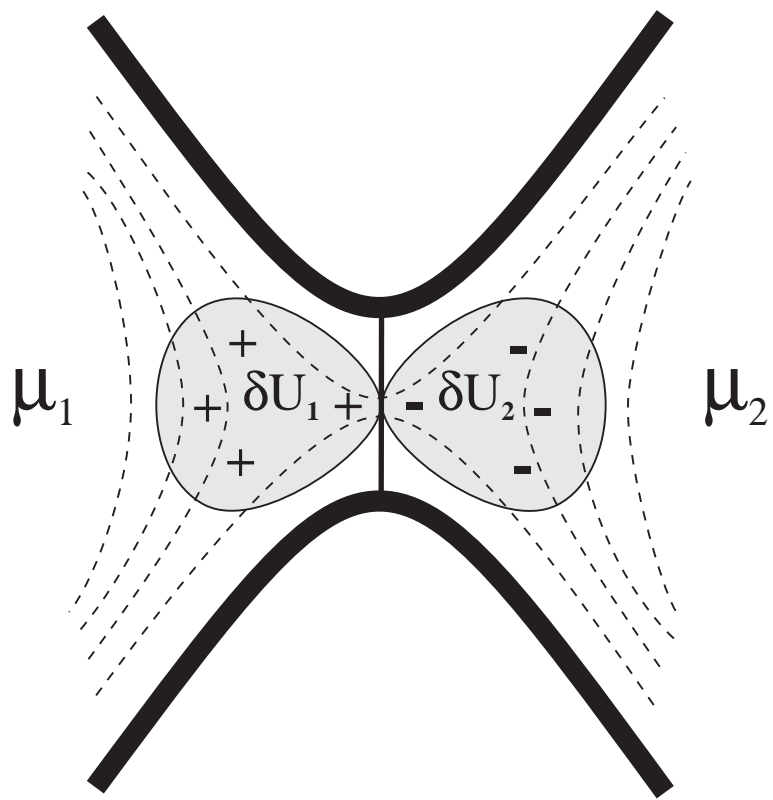
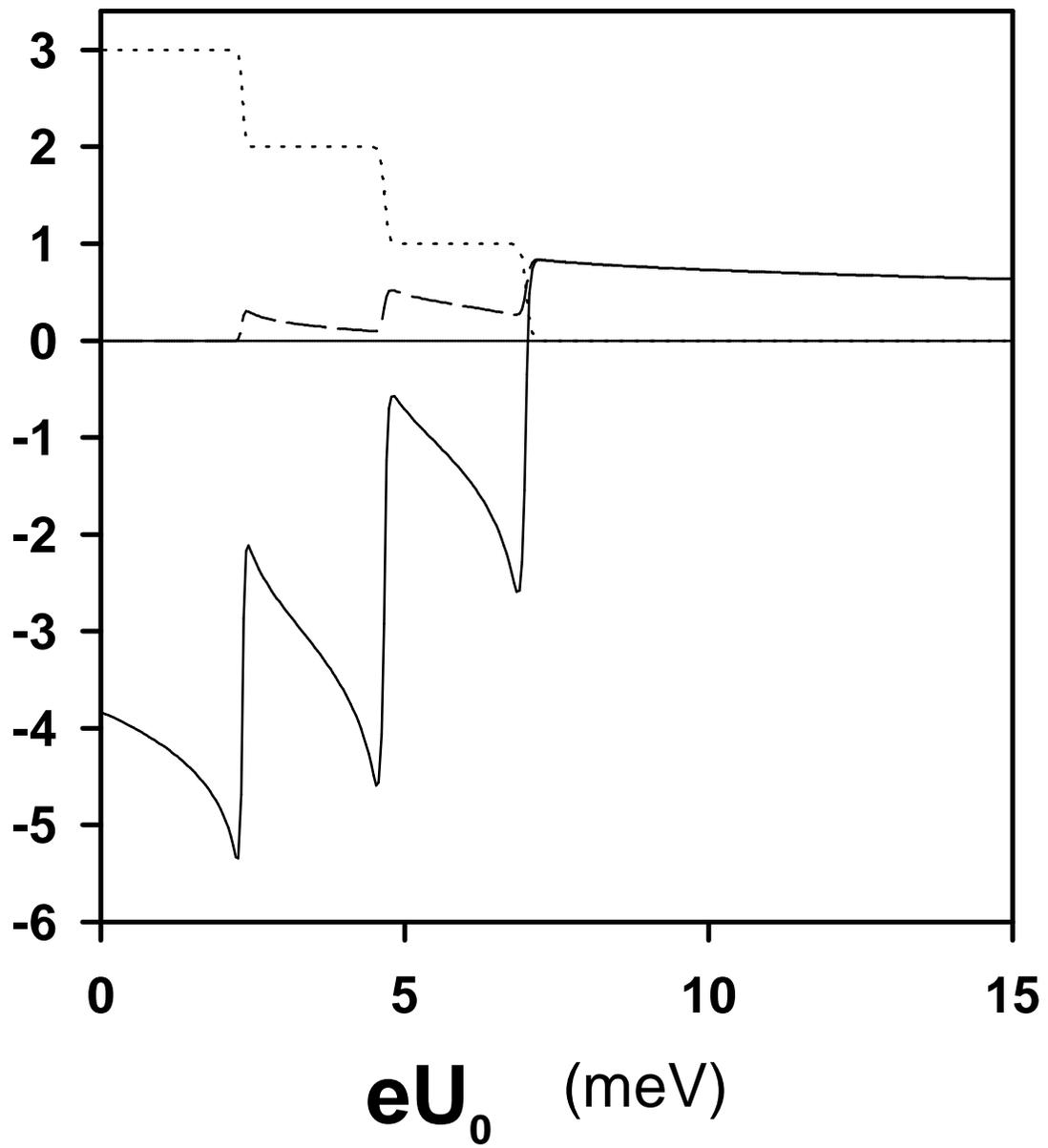
\end{document}